\documentclass[showpacs,showkeys,prc,twocolumn]{revtex4}
\usepackage{graphicx}
\usepackage{amsmath,amsfonts,amssymb}
\usepackage{natbib}

\usepackage{color}

\begin{document}

%\title{An Apparatus for the Investigation of Ultracold Neutron Production under Applied Magnetic Field}
\title{Ultracold Neutron Production in a Pulsed Neutron Beam Line}
\date{\today}
%\date

\author{C.M. Lavelle}
%\email{chlavell@indiana.edu}
\author{W. Fox}
\author{G. Manus}
\author{P.M. McChesney}
\author{D.J. Salvat}
\author{Y. Shin\footnote{currently at Department of Physics, Yale University, New Haven, CT 06511}}
\author{C.-Y. Liu}
\email{cl21@indiana.edu}
\affiliation{Physics Department, Indiana University, Bloomington, IN 47408}
\author{M. Makela}
\author{C. Morris}
\author{A. Saunders}
\affiliation{Physics Division, P25, Los Alamos National Laboratory, Los Alamos, NM 87544}
\author{A. Couture}
\affiliation{LANSCE Division, NS, Los Alamos National Laboratory, Los Alamos, NM 87544}

\author{A.R. Young}
\affiliation{Physics Department, North Carolina State University, Raleigh, NC 27695}

\begin{abstract}
We present the results of an Ultracold neutron (UCN) production experiment in a pulsed neutron beam line at the Los Alamos Neutron Scattering Center. The experimental apparatus allows for a comprehensive set of measurements of UCN production as a function of target temperature, incident neutron energy, target volume, and applied magnetic field. However, the low counting statistics of the UCN signal expected can be overwhelmed by the large background associated with the scattering of the primary cold neutron flux that is required for UCN production. We have developed a background subtraction technique that takes advantage of the very different time-of-flight profiles between the UCN and the cold neutrons, in the pulsed beam. Using the unique timing structure, we can reliably extract the UCN signal. Solid ortho-D$_2$ is used to calibrate UCN transmission through the apparatus, which is designed primarily for studies of UCN production in solid O$_2$. In addition to setting the overall detection efficiency in the apparatus, UCN production data using solid D$_2$ suggest that the UCN upscattering cross-section is smaller than previous estimates, indicating the deficiency of the incoherent approximation widely used to estimate inelastic cross-sections in the thermal and cold regimes.  
%Solid oxygen is a candidate material for Ultracold Neutron (UCN) production due to its low neutron absorption and magnetic excitations, the latter of which may provide additional modes for enhanced UCN production. The contribution of magnetic modes to UCN production can be probed by applying magnetic fields to perturb the dispersion relation of the magnetic excitations, altering the UCN production cross-section.  We have built an apparatus to investigate solid oxygen as a potential next generation UCN source which incorporates an applied magnetic field (up to 5.5 T).  We describe the calibration of this apparatus using a solid ortho-deuterium UCN production source.  We also discuss a novel approach for the elimination of backgrounds associated with cold neutrons scattered from within a short, large area UCN guide.
\end{abstract}



\maketitle

\section{\label{sec:intro}Introduction}

%Nano-eV neutrons are free neutrons with equivalent thermal energy on the order of 1 mK.  For this reason, they are
%termed Ultra-Cold Neutrons (UCN).  These unique particles find application to fundamental physics because their energy is sufficiently low that they may be confined in material or magnetic traps, which allowing for precision spin manipulations and clear detection of decay products.  For example, measurements of the neutron electric dipole moment and the neutron lifetime which use bottled UCN can test physics beyond the standard model, and are complimentary to cold neutron (meV) beam experiments \cite{golub1991,pend2000,abele2008}.

Ultracold Neutrons (UCN) are free neutrons with maximum energy of 300~neV, equivalent to $\sim$ 1 mK.  Their kinetic energy is so low that they may be contained and accumulated in material bottles, magnetic traps, and the Earth's gravitational field, for durations up to hundreds of seconds \cite{ageron1985}. 
The kinetic energy is comparable to the Zeeman splitting in a magnetic field of a few Tesla, and neutron state with 100\% spin polarization can be prepared through simple field filtering. 
Storability and easy polarization of UCN make them the tool of choice in many experiments to measure fundamental properties of the neutron to unprecedented precision.\cite{golub1991,abele2008} 
The most precise measurements of the neutron electric dipole moment \cite{edmbaker2006} and the neutron beta-decay lifetime \cite{Serebrov2008,huffman2000} all employ UCN in a trap. Any innovation that increases the storable UCN density and the deliverable UCN flux will simplify many technical difficulties in these challenging experiments.
For example, a larger UCN density would reduce the size of most EDM experiments, making the implementation of stringent magnetic field uniformity and large electric field a less daunting task. In addition to applying UCN for the studies of fundamental physics, there are also many applications to condensed matter physics and possible enhanced sensitivities derived from the long wavelength of UCN\cite{golub1996}, if a more intense source were to become available.

%The most often utilized method of UCN production is single-event downscattering, first proposed by Golub and Pendlebury \cite{golub1975}.  Incident cold neutrons lose kinetic energy to become UCN by exciting collective excitations (usually a single phonon) in the interacting medium. Any medium possessing strong scattering and small nuclear absorption, regardless of the excitation's details (phonon, roton, magnon, libron, etc.), could potentially make a good UCN converter.  The only requirement is that the excitation's dispersion intersect the free neutron dispersion relation, $\hbar \omega = \frac{\hbar^2 k^2}{2m_n}$. The reverse process process of energy gain from the medium, or up-scattering, is suppressed via the Boltzmann factor by lowering the temperature.  Successful UCN sources of this type utilize superfluid helium-II \cite{huffman2000} and solid deuterium \cite{saunders2004},  which dissipate neutron energy through phonon emission.

UCN are already present in the Maxwell-Boltzmann spectrum of thermalized neutrons emerging from fission reactors. 
However, the percentage of the low-energy population is so low that even the most powerful research reactors cannot easily deliver high enough UCN density to be of interest. 
The use of a cold neutron moderator could shift the energy distribution to as low as 30 K, resulting in an increased UCN flux. 
In addition, many tricks have been implemented, including gravitational deceleration \cite{Kosvintsev1977} and Bragg deflection on mechanical turbines to further slow down neutrons \cite{steyerl1975}. 
To increase the phase-space density beyond the limit imposed by the Liouville theorem, the most efficient method is to dissipate the neutron energy through excitations in condensed matter. Phonons in many materials have energies ($\sim$ meV) comparable to that of moderated cold neutrons. 
Golub and Pendlebury \cite{golub1975} first proposed a ``superthermal'' source to use superfluid helium as a UCN converter. In this
type of source the neutron gives up all of its kinetic energy by exciting collective excitations in the interacting material. 
The depletion of the UCN population via absorption of the same excitation energy (so-called ``upscattering'') can be suppressed to an arbitrarily small level simply by lowering the temperature of the converter.
In the superthermal source, the number density of UCN could accumulate for durations as long as the $\beta$-decay lifetime of free neutrons.

To preserve a sizable population of UCN ($\sim$ mK) in a source environment of a few Kelvins, it is essential to avoid re-thermalization by delaying upscattering. It is also important to control sources that could lead to UCN loss, such as nuclear absorption. Any medium possessing small nuclear absorption and significant neutron scattering cross-sections, regardless of the details of the excitation, are potential candidates. 
Successful implementation of this idea have been demonstrated in superfluid helium~\cite{golub1983,kilvington1987,masuda2002} and solid deuterium~\cite{morris2002,saunders2004,atchison2005,Gutsmiedl2009}. Both of these UCN converters have phonon excitations that match well with the incident cold neutrons for efficient energy transfer through single inelastic scattering. 
Other materials could potentially make even more efficient UCN converters. For one, $^{16}$O has a neutron absorption cross-section five times smaller than that of $^2$H. In addition to phonons, solid oxygen (s-O$_2$) in its low temperature phases has strong magnetic interactions which could be harnessed for UCN production \cite{liu_thesis}.  
These magnetic excitations have been widely studied using Raman scattering\cite{Kreutz2003,Medvedev2003} as well as neutron scattering \cite{stephens1986,dunstetter1996,Bermejo1998}. It has been shown that the low temperature phases possess very different dynamics that often entangle translational, librational, and spin excitations through the orientation-dependent couplings between the diatomic oxygen molecules on the solid lattice sites. 

We present an apparatus used to test s-O$_2$ as a UCN converter. A background subtraction technique has been developed to enable collection of UCN production data in a pulsed neutron facility, where the cold neutron background can be significant. 
To extract the absolute efficiency of UCN production from the source, we carefully characterize the efficiencies of UCN transmission and UCN detection.  We calibrate the efficiency of the apparatus using the well-studied solid ortho-deuterium (o-D$_2$) as a standard reference.  Here, we present a comprehensive test of the physics of UCN production in o-D$_2$ using this apparatus.
The experimental results of UCN production in s-O$_2$ (including dependence on incident neutron energy, source volume, and the applied external magnetic fields) will be presented in forthcoming papers.  
%In this paper, we limit our discussion to the understanding of the performance of this apparatus using the ortho-deuterium solid (o-D$_2$).

\section{\label{sec:apparatus}Experimental Methods}
\subsection{Overview}

The experiment is carried out on a cold neutron beam line, where the cold neutron flux and energy spectrum are well characterized. The neutron flux is collimated and monitored continuously. Even though the available flux is significantly smaller compared to a previous experiment using a dedicated mini-spallation target \cite{saunders2004,masuda2002}, the removal of the uncertainties associated with the production and transport of the cold neutron flux is essential in understanding the results of UCN production. The new apparatus consists of a neutron scattering target cell which allows us to grow solid target in-situ. The cell is placed in a cold neutron beam line to act as the UCN production source.

UCN, once produced inside the source, are extracted and then detected in a neutron detector placed sufficiently far away from the primary neutron beam. The strategic placement of the detector helps to control the 
backgrounds.
Elastically scattered neutrons from the intense cold neutron beam are the dominating source of background in this experiment. Determining the UCN extraction and transport efficiency is the major challenge in determining the absolute UCN production cross-section.  We therefore employ an o-D$_2$ converter to benchmark the performance of the apparatus. Production of UCN in o-D$_2$ has been demonstrated at PNPI \cite{serebrov2000a} and the superthermal behavior was measured at LANSCE \cite{morris2002,saunders2004}. Further neutron transport and UCN production cross-sections in o-D$_2$ have been carefully carried out at PSI \cite{atchison2005,atchison2007} and by the Mainz/Munich group \cite{Gutsmiedl2009}.

\begin{figure}[!b]
\centering
\includegraphics[width=3.5 in]{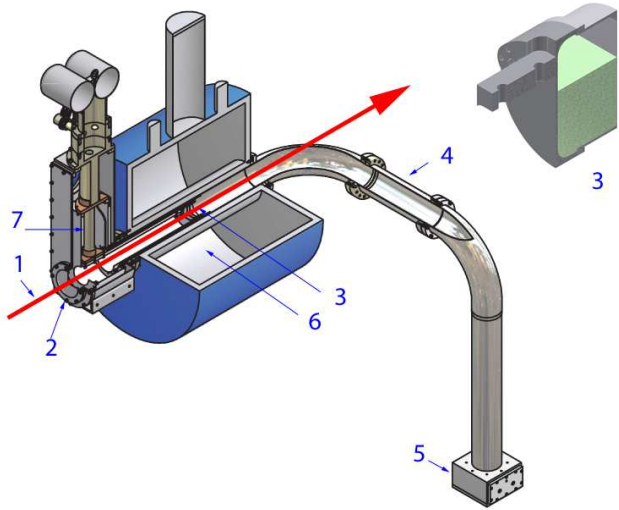}
\caption{\label{fig:app}A schematic of the apparatus. Cold neutrons (1) enter the front Aluminum window (2) and are incident on the cryogenic target cell (3).  Ultra cold neutrons (UCN) produced within the cell diffuse out of the cell and into a highly polished stainless steel UCN guide (4), which direct them through two 90 degree bends into a UCN detector (5). The cell itself resides a  5.5 T (maximum) superconducting solenoid, housed inside a liquid helium cryostat (6). The target cell is cooled by a pulse tube refrigerator (7). A cut-away view of a partially filed cell (3) is shown in the inset.}
\end{figure}

A unique feature of the apparatus (illustrated in Fig.~\ref{fig:app}) is the integration of a high magnetic field, which allows for the study of the magnetic excitations in s-O$_2$. The target cell is placed in the center of a superconducting (SC) solenoid with the axis aligned along the direction of the neutron beam. A cylindrical stainless steel guide serves as the UCN transport guide, as well as the insulating vacuum for the cryogenic target cell. The guide extends beyond the warm bore of the SC solenoid and connects to the rest of the UCN guide system. 

The cell is larger than the cold neutron mean free path (MFP) in o-D$_2$ so that the majority of incident cold neutrons scatter at least once before escaping the cell. Inside the filled cell, a small fraction of these scattering events, less than 1 part in $10^6$, results in the production of a UCN. Some of the produced UCN escape from the cell, enter the guide system, and move toward the detector.

The UCN guide system was made of stainless steel tubes 10.16 cm in diameter. The pharmaceutical-grade guides are internally electro-polished to about $\sim0.25$ $\mu$m mean variation of the surface roughness. The total length was approximately 2 m with two 90-degree bends to direct the UCN perpendicularly out of the primary cold neutron beam, then vertically downward into the detector. The target cell is cylindrical, with the cold neutron beam impinging on the front surface and the UCN primarily extracted from the opposite surface downstream.  

%The final UCN yield depends strongly on both the cold neutron attenuation and the UCN MFP in the converter material.  Our Monte-Carlo simulation estimates that only about 2\% of all UCN produced in the cell are detected (see Section \ref{sec:MC}).

Ostensibly, only UCN and very cold neutrons (VCN), the latter with a lower probability, could totally reflect from the internal walls of the guide system and 
propagate to the detector.  
However, we found a significant background due to cold neutrons which are diffracted and/or scattered incoherently by the section of UCN guide that intersects the beam. Even though the probability is quite low ($\sim 3\times10^{-8}$), the large flux of cold neutrons passing through the apparatus leads to a background comparable to the UCN count rate. 
In sec~\ref{sec:signal}, we will show that this background
can be removed by using its very distinct timing structure. With the relatively short guide system, the apparatus has a high UCN throughput, without being plagued by the cold neutron background.

Eventual use of s-O$_2$ as a UCN converter requires good temperature control of the cryogenic target cell, essential to reproducing the thermodynamic conditions for repeatable crystal growth. The major challenge arises from the reduction of the molar volume of s-O$_2$ by 12.5\% upon cooling through the three solid phases ($\gamma, \beta, \alpha$) at saturated vapor pressure. A discrete 5\% change at the $\gamma-\beta$ transition \cite{defotis1981,jez1993} at 44 K is the bottleneck for attaining large-sized s-O$_2$ cryocrystals in the $\beta$ and $\alpha$ phases. Poor crystal quality, which adversely affects UCN extraction, needs to be controlled in order to gain understanding of the physics of UCN production in s-O$_2$. Density fluctuations (caused by cracks) over a range comparable to and/or larger than the UCN wavelength could result in additional scattering and lead to a reduced MFP \cite{golub1991}. Several groups have observed this effect in s-D$_2$. In particular, Atchison and coworkers \cite{atchison2005} measured a $\sim$20 barn increase of the total cross-section in o-D$_2$ after several thermal cycles. There are no data on s-O$_2$.

Unlike o-D$_2$, the MFP of UCN in s-O$_2$ is not limited by the incoherent scattering length, and is theoretically infinite because of the zero nuclear spin of the oxygen nucleus $^8$O. On the other hand, any additional voids and cracks could limit the MFP of UCN and significantly alter the extraction efficiency. With this apparatus, we have investigated different factors limiting the MFP of UCN in the production target by changing the length of the target cell along the cold neutron beam axis.

\subsection{\label{sec:CN}Cold Neutron Beam}

The UCN production apparatus was constructed and tested at the Indiana University Cyclotron Facility, and then installed on the Flight Path 12 (FP12) in the Lujan Center at LANSCE in August 2008. This neutron facility generates pulsed neutrons at 20 Hz from a spallation target. FP12 is coupled to a liquid hydrogen neutron moderator with straight neutron guides. Details of the neutron source emission time distribution, guide performance, and overall intensity can be found in \cite{russina2003, fuzi2006, seppo2008}. The neutron spectrum peaks at 3.3~meV ($\sim$40~K) \cite{seo2004,ooi2006}. The neutron energy is determined from the time-of-flight (TOF), $t$, over the flight path length, $L$, by
\begin{equation}
E(t)=\frac{1}{2} m L^2 t^{-2}
\end{equation}
with the corresponding energy resolution ($\delta E$) of
\begin{equation}
\left(\frac{\delta E}{E}\right)^2 \approx 4 \left(\frac{\delta t}{t}\right)^2+4\left(\frac{\delta L}{L}\right)^2.
\end{equation}
Here $\delta t$ is the emission time, and $\delta L$ is the uncertainty in the flight path length.
The arrival of the proton pulse defines the $t_0$, however, details of moderator geometry, neutron slowing-down and diffusion within the neutron source itself limit the timing resolution to the emission time \cite{russina2003}. The emission time is the time spread for neutrons of a given energy.
The fastest neutrons observed by TOF are around 100~meV, with a resolution of around 2\% for $\delta t \sim$ 150  $\mu$s and $L=$21.1 m. The relative uncertainty in the flight path length is negligible.
A frame overlap chopper absorbs the long wavelength neutrons and restricts the lower limit of the incident neutron energy to E$>$1.2~meV. The most recent measurement \cite{seppo2008} reported the neutron flux at the sample position to be $(2.0\pm0.1)\times 10^7$ n~cm$^{-2}$s$^{-1}$, integrated over the spectrum from 1.2 to 20~meV, for 100~$\mu A$ proton current on the spallation target. This measured intensity includes a boost in flux from an m=3 guide.

\begin{figure}[!t]
\centering
\includegraphics[width=3.5 in]{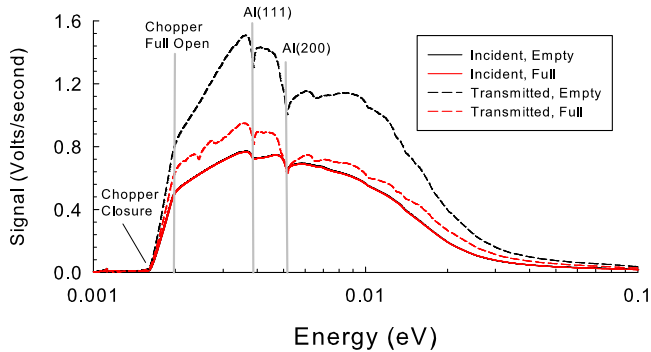}
\caption{\label{fig:raw_mon} An example of the neutron monitor signals $I(t)$ recorded with the waveform digitizer with an averaged proton current of 100~$\mu$A on the spallation target.  The empty cell signal (black) is shown together with the 4 K o-D$_2$ filled cell data (red) for both the incident beam monitor (M$_1$) and the transmitted beam monitor (M$_2$.)   The solid gray lines show the features which are aligned to determine the monitor position relative to the incident beam monitor, which is fixed at 21.11 m.  Vertical axis is the voltage output from the amplifier, and its scale is set by gain resistor in the amplifier circuit. The monitor signal $I(t)$ is the integrated voltage signal per time channel normalized to the duration of data-taking, i.e., 1000~s.}
\end{figure}

Typical neutron spectra collected with our instrument on FP12 are shown in Fig.~\ref{fig:raw_mon}, using an empty cell and a cell filled with o-D$_2$ at 4.81$\pm$0.02 K. The charge-integrated voltage signals from the cold neutron monitors are digitized with a 12-bit waveform digitizer with a sample rate of 102.4 kHz. 
The neutron intensity $I(t)$ is normalized to the average proton beam current of 100~$\mu$A. A few features in the beam line modifies the expected Boltzmann spectrum of neutrons from the moderator. First, the signal was zero at low energies because the frame-overlap chopper was closed to prevent the slower neutrons from appearing in the next frame.  The neutron flux increased as the chopper was opened, leading to a rising edge in the spectrum. The two most pronounced dips in the spectrum were the Al [111] and Al [200] Bragg edges from aluminum structural elements and vacuum windows in the cold neutron beam path. Additional peaks due to Bragg scattering from solid D$_2$ and the stainless steel UCN guides in our apparatus appeared in the downstream monitor (M$_2$).

The neutron flux monitor M$_1$ was mounted at the end of the cold neutron guide, and was kept at a fixed position of 21.11$\pm$0.03 m downstream from the moderator. By aligning the Bragg edges, we calibrated the position of M$_2$ to be 22.75$\pm$0.04 m from the moderator, with the error bar on the distance determined from the width of the Bragg peak. Neutron signals measured by M$_1$ were unchanged between runs, indicating good stability.

The flux of transmitted cold neutrons (collected in M$_2$) is used to calculate the total scattering cross-section of the target material (o-D$_2$) during calibration runs.  
%The deduced neutron cross-section test our understanding of the apparatus.  
Several factors related to the non-ideal geometry need to be taken into account in calculating the total cross-sections. To avoid blocking the fill/vent line below solidification temperature and thus prevent pressure hazards, the cell was only partially filled with liquid (volume fraction $f=0.601\pm 0.014$  of o-D$_2$). The fill volume was determined from the pressure drop in the storage volume, together with the known density of the liquid.  With the cell uniformly illuminated, the neutron flux measured in M$_2$ thus consists of both un-attenuated neutrons from the upper part of the cell and the attenuated neutrons from the lower part of the cell.
We determine the cross-section using the following algorithm:
\begin{equation}
n \sigma(E)x=\ln\left(\frac{f}{\frac{I_{filled}}{I_{empty}}-(1-f)}\right) \label{eqn:totalxs},
\end{equation}
where $\sigma$ is the total neutron cross-section per molecule, $n$ is the molecular number density, and $x$ is the target thickness (3.56 cm). The combined effects of digitizer voltage resolution and frame overlap chopper restrict our measurement in energy range from 1.2 to 30~meV.

\begin{figure}[!t]
\includegraphics[width=3.5in]{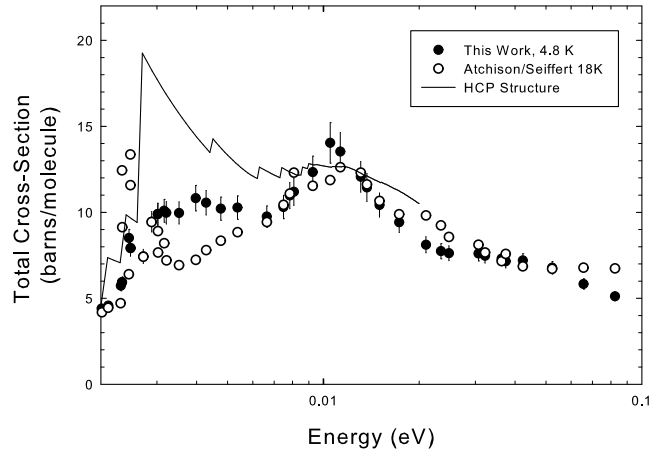}
\caption{\label{fig:totalxs}The total cross-section of o-D$_2$ derived from Eq.(\ref{eqn:totalxs}) is compared with measurement done by \protect\cite{atchison2005a}. The theoretical cross-section (solid line) assuming a hexagonal closed packed (HCP) structure is shown for comparison. }
\end{figure}

The total cross-section for o-D$_2$ is shown in Fig.~\ref{fig:totalxs}, together with the data of Seiffert \cite{seiffert1970}. We obtain the best agreement with the previous measurements by adjusting the fill fraction to $f=0.57\pm0.02$. This reduced value reflects the expected 4~\% fractional change of volume upon cooling o-D$_2$ to 4~K \cite{manzhelii98}. However, this volume change could also represent a 1.6 $\pm$ 0.7~mm thick layer of solid frozen to the walls in the unfilled, upper portion of the cell.  Given the high vapor pressure of o-D$_2$, both effects likely contribute to the filling fraction $f$ smaller than that estimated by the pressure drop record. Finally, we note the cell is about 1.6 MFP thick, and thus multiple scattering was likely to occur, leading to an excess in the measured cross-section as observed in the range below 10~meV.

\subsection{Gas Handling System and Cryogenic Target Cell}

\begin{figure}[!b]
\includegraphics[width=3.5in]{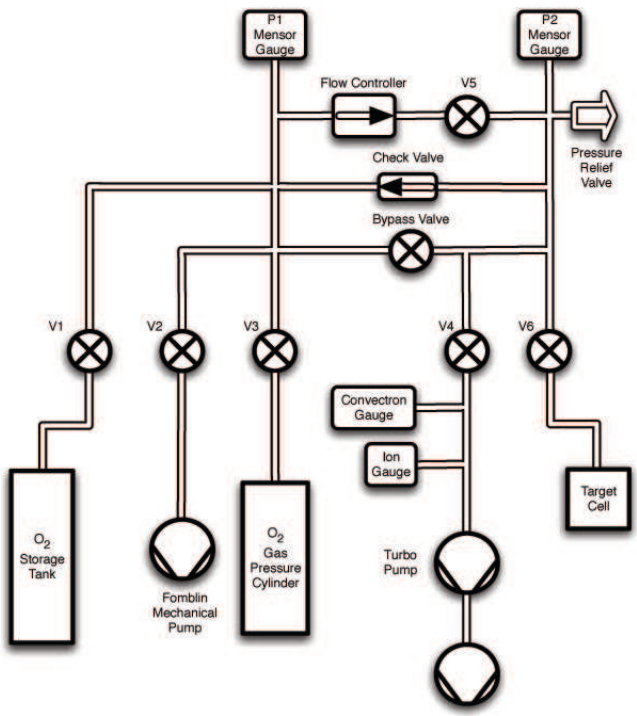}
\caption{\label{fig:GHS} Schematic of the gas handling system (GHS).}
\end{figure}

The materials of interest to this experiment, o-D$_2$ and O$_2$ gas, are both highly volatile, and thus much care was put into the design and the construction of the Gas Handling System (GHS). A schematic of GHS is shown in Fig.~\ref{fig:GHS}.
The GHS was constructed out of all stainless steel VCR tubing and connectors with all-metal seals. Only dry pumps and fomblin pumps were connected to the GHS. In the beginning of each run, it was cleaned by pumping until the pressures were less than 10$^{-4}$ Pa measured on the GHS panel before any gas was introduced. 
 The cell was filled from a 200~l storage tank via a flow controller on the GHS.  A check valve was installed between the cell and the storage tank. It was set to open at a $1.3\times10^4$ Pa pressure differential to relieve overpressure in the cell back to the storage tank to contain the volatile gas. In addition, a drop-out plate on the UCN guide system (which contains the cell) mitigated any catastrophic overpressure in the event the cell should burst.

There was only one fill line (0.3175 cm diameter SS tubing) that connected the GHS to the cell. In order to prevent blockage, the fill line had to be kept at temperatures above the solidification temperature of the material under study. In addition, the fill line was connected to the cryogenic cell using a stainless steel flange. The low heat conductivity of the stainless steel flange allowed for a large temperature gradient, isolating the cryogenic cell from the warm fill line. The fill line was wrapped with a nichrome wire heater and temperature-controlled with a PID loop.

Inside the cell, the solid was frozen from the liquid phase. The triple points are 18.7 K, 16.9 kPa for D$_2$ and 54.4 K, 152 Pa for O$_2$. The o-D$_2$ gas was borrowed from the dedicated UCN source for the UCNA experiment \cite{ucna2009}. The ortho-para ratio of the gaseous D$_2$ was measured via Raman spectroscopy at the beginning and the end of the experimental run. The initial measurements showed a contamination of 2.9 $\pm$ 0.02~\% para-D$_2$ and 0.17 $\pm$ 0.11~\% hydrogen deuteride (HD). The para fraction increased 0.2~\% over the 50-day duration of the experiment, and the HD impurity level remained constant within error. After the D$_2$ calibration run, the D$_2$ gas was thoroughly pumped out before the oxygen gas was introduced. O$_2$ was supplied via a high pressure gas cylinder at 99.999\% purity.

The target cell consists of a type 6061 Al cylindrical cup (several lengths available) bolted to a circular base flange, and installed horizontally in the magnet bore. The cell wall downstream of the beam direction consists of a thin Al window (0.254 mm) to facilitate UCN transmission and subsequent detection.   
To locate the cell within the center of the magnetic solenoid, the top flange is attached to one end of a high purity (99.999\%) Al bar $\sim$30 cm in length. The Al bar is cooled by a pulse tube refrigerator with a total cooling power of 1.5 W at 4 K. The target cell is made vacuum-tight with an indium wire seal. Throughout the experiment, three different cell lengths were used with the base flange/cup combinations shown in table \ref{tbl:cells}. The inner diameter of the cup side of the cell is 6.4 cm, and the base flange side is 6.67 cm. We used the medium cell for the o-D$_2$ calibration runs. 

% a small base, 1.143 cm long, with a flat plate cover, total volume 39.82 cc; and the same small base with a cup to extend the total length and volume to 3.556 cm and 117.4 cc.  Finally, the largest cell employs a long base with the previous cap for a total length of 8.611 cm and volume of 290.2 cc. %

\begin{table}
\caption{\label{tbl:cells} Cell geometries used in the experiment (see inset in Fig.\ref{fig:app})}
\begin{tabular}{|c|c|c|c|}
\hline \hline
Cell Type & Base Length (cm) & Cup Length (cm) & Volume (cc) \\ \hline
Small & 1.143 & 0 (flat window) & 38.82 \\
Medium & 1.143 & 2.413 & 117.4 \\
Large & 6.198 & 2.413 & 290.2 \\
\hline \hline
\end{tabular}
\end{table}

The cell was surrounded by two layers of cold shields. The inner cold shield was connected to the cold head (cold stage) of the pulse tube refrigerator, while the outer shield was cooled by the warm stage ($\sim$ 50 K). The whole assembly was placed inside the stainless steel UCN guide, which runs through the warm bore of a separate helium cryostat that houses the superconducting magnet. The inner cold shield was lined with a thin nickel foil (50 $\mu$m) to increase the collection efficiency of UCN emerging from the sides of the cell.  Several G10 rings were used to center the cold shields and the cell, preventing thermal shorts between the cold shields and the inner wall of the UCN guide. Stable temperature operation with fluctuations no larger than 3 mK was achieved from 4.8 to 80 K with an empty cell.  When filled, fluctuations of each temperature sensor increased to 15-20 mK. In addition, the temperature gradient across the cell when filled with solid o-D$_2$ increased to $\sim$ 60 mK.
The UCN guide was evacuated to less than 10$^{-4}$ Pa to provide insulation required to operate the cryogenic target cell. Throughout the experiment, a residual gas analyzer constantly assayed the gas composition in the guide vacuum, to monitor for leakage from the cell.

We used Lakeshore CERNOX 1050-SD Resistive Temperature Detectors (RTDs) for thermometry because of their resistance to high radiation and insensitivity to high magnetic fields. Temperature readout and heater power was controlled using a Lakeshore 330 Temperature Controller. We developed a MATLAB based DAQ to implement slow controls for the RTDs, heaters, and pressure sensors. The same DAQ also controlled a fast digitizer to read out the cold neutron monitors, and the multi-channel scalar (MCS) used to collect TOF spectrum from the UCN detector.  This DAQ program thus provided a centralized platform for thermometry control and monitoring, data acquisition, visualization, as well as the on-line data analysis.

\section{Results}
\subsection{\label{sec:UCNPHS}Ultracold Neutron Detection}
An improved version of the multi-wire proportional counter \cite{morris_det} was used to detect UCN. Extensive neutron shielding using 0.5~m thick borated polyethylene and B$_4$C powders surrounding the detector was essential to reduce the background rate from 300 Hz (un-shielded) to 134 $\pm$ 2 mHz. The counter, filled with 1 kPa of $^3$He gas, has a small, yet non-zero efficiency of 1.3~\% to detect cold neutrons with a 40~K Maxwellian spectrum. A typical UCN detector count rate with a 60\% filled cell of o-D$_2$ at 5 K with 100~$\mu$A proton current is 612 $\pm$ 13 mHz. The beam-off ambient neutron backgrounds were only 20 mHz, and thus most of the 134 mHz observed is due to the transmitted cold neutrons which were elastically scattered in the UCN guide. 

Inside the FP12 cave, there was a high level of $\gamma$ radiation as a result of neutron captures.  The $\gamma$s were detected by the UCN detector as low energy pulses, but their intensity was high enough to prevent clean separation between the $\gamma$ background from the higher energy neutron peak. We set the counter threshold to read the full energy neutron peak.  The large threshold enhanced $\gamma$ rejection to nearly 100\%, at a cost of reducing the efficiency for neutron detection to  85 $\pm$ 3~\%. Note that the detector front window also attenuates the UCN flux, the effect of which will be included as a part of the transmission efficiency.
%\footnote{We compared the resulting spectrum to the pulse height spectrum presented in \cite{morris_det} and found the overall counting efficiency reduced somewhat to 85$\pm$3 \% of the neutron induced pulses.}

\begin{figure}
\centering
\includegraphics[width=3.5in]{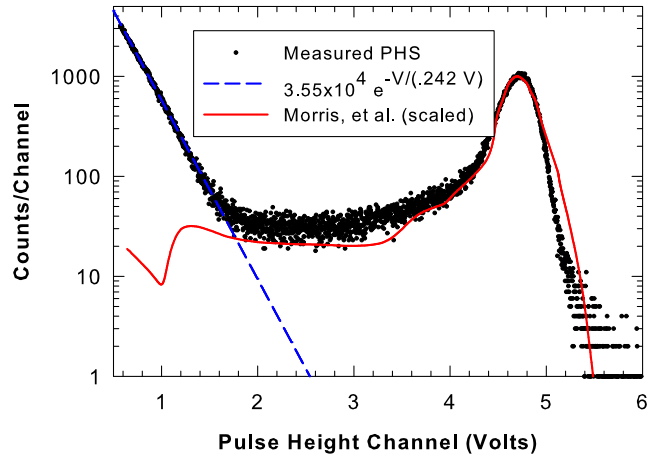}
\caption{\label{fig:ucn_phs} The pulse height spectrum measured by the multi-wire proportional counter. Solid dots are the measured data. The red curve is the theoretical fit of the spectrum including the wall effects from the daughter nuclei of proton and triton. The solid blue line is the fitted gamma background. With the threshold set to enclose only the full energy peak, the UCN detecting efficiency is 85 $\pm$ 3~\%. }
\end{figure}

\begin{figure}[!b]
\centering
\includegraphics[width=3.5 in]{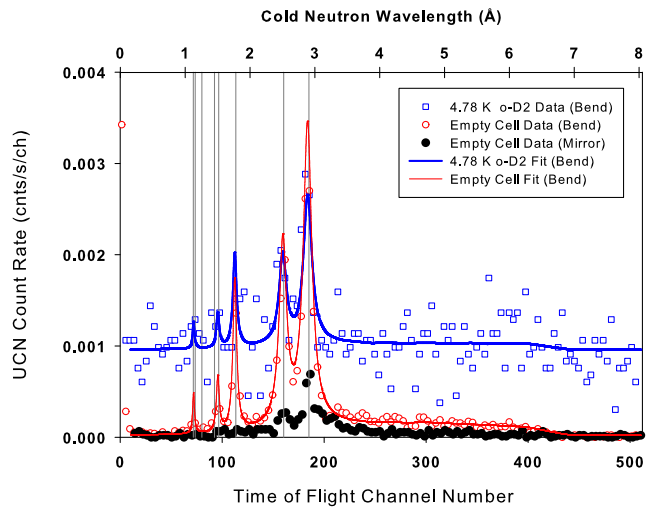}
\caption{\label{fig:bkgfit} A MCA spectrum of the background collected with empty cell (red circles) is fitted to Eq.(\ref{eqn:fit}) (red curve). The grey vertical lines indicate locations of the first 7 Bragg diffraction peaks from the stainless steel guide. When a mirror is used (black solid circles), the CN Bragg peaks drop in intensity, with some residual background scattered from the mirror's support frame and the vacuum chamber. UCN signal spectrum collected from the 4.8 K o-D$_2$ target is plotted (blue squares) and fitted to Eq.(\ref{eqn:fit}) (blue curve). Note the relative intensities of Bragg peaks of the filled and the empty cell. The reduced Bragg peak in the filled cell is a result of the attenuation of cold neutrons by the presence of o-D$_2$ solid.}
\end{figure}

Signals from the UCN detector were recorded using a multi-channel scalar (MCS), triggered by the proton pulse on the spallation target at 20 Hz. The UCN counts were recorded for 50 ms into the time channels on the MCS. 
To collect enough data for statistical analysis, TOF spectra are accumulated by the MCS for 1000 s for each set of experimental condition. Without the loss of proton pulses, 20,000 passes of spectrum are collected for each run. 
The first 1.2 ms of the detector response was dominated by the initial burst of radiation from the proton pulse, and is cut in the analysis. The TOF of UCN to travel from the production cell to the detector is much longer ($\sim$several seconds) than the time between cold neutron pulses (=50~ms), so we expect the UCN signal to have a uniform TOF spectrum. On the other hand, the measured TOF spectrum displays several prominent peaks, as shown in Fig.~\ref{fig:bkgfit}. These peaks originate from the Bragg diffraction of cold neutrons from the stainless steel UCN guide that intersects the cold neutron beam. The TOF information allows one to associate the observed peak to the Bragg diffraction peaks. The width of the peaks is due to the large angular acceptance of the UCN guide system.  The guide was made of 316L stainless steel 3.175~mm thick.  Its lattice structure is face centered cubic with lattice parameter 3.6 $\AA$ \cite{Steuwer2003}. Using a total flight path to the detector of 24.5~m, and scattering through an arc of 44$^{\circ}$, we reproduced these peaks at the correct positions in the TOF spectrum.  The expected and measured peak positions are summarized in Table \ref{tbl:bragg}.
In addition to these distinct peaks, there exists a long TOF tail that resembles the spectral shape of the incident cold neutron spectrum. We attribute the long TOF tail to the incoherent elastic scattering of neutrons from the UCN guide.
\begin{table}
\caption{\label{tbl:bragg}The expected and observed locations of the Bragg reflections in the background spectrum for an FCC lattice, a=$3.6\AA$. After re-binning, measurement wavelength resolution is $\sim0.05 \AA$. The intensities of the [222] and [400] peaks were too low to be observed.}
\begin{tabular}{|c|c|c|}\hline\hline
\textbf{Indices}&\textbf{Expected Peak $(\AA)$}&\textbf{Fitted Peak $(\AA)$}\\\hline
1 1 1 & 2.89 & 2.90\\
2 0 0 & 2.50 & 2.52\\
2 2 0 & 1.77 & 1.78\\
3 1 1 & 1.51 & 1.51\\
2 2 2 & 1.44 &  - \\
4 0 0 & 1.25 &  -  \\
4 2 0 & 1.12 & 1.14\\\hline\hline
\end{tabular}
\end{table}

In order to further reduce this source of background, we replaced the stainless steel bend with a 45$^{\circ}$ mirror made of a thin nickel foil only 0.127 mm thick, housed inside a stainless steel tee . This reduced the amount of background CN by an additional factor of 4 (from 134~mHz to $34.8 \pm 1.3$~mHz), but it also reduced the UCN transmission efficiency to 73 $\pm$ 2~\% of the previous configuration.  Nevertheless, the signal to background ratio improved by more than a factor of two. The background TOF spectrum with the mirror is compared to that with the bend in Fig.~\ref{fig:bkgfit}.

\section{\label{sec:signal}Background Subtraction}

%To collect enough data for statistical analysis, TOF spectra are accumulated by the MCS for 1000 s for each set of experimental condition. Without the loss of proton pulses, 20,000 passes of spectrum are collected for each run. 
%Each spectrum is triggered by a proton pulse, and each pulse is separated by 50 ms in time from the next identical pulse.
For the data obtained in the present geometry, a two parameter fit to the CN-induced background and UCN signal is implemented.
First, the model used to fit the raw TOF spectrum is the following:
\begin{eqnarray}
C(t)&=& C_{ucn} + C_{bg}(t) \nonumber \\
&=& C_{ucn}+(B_{CN}(t)+C_B) \label{eqn:sig}
\end{eqnarray}
with
\begin{equation}
B_{CN}(t) = \sum_{i=1}^{5}{A^{coh}_i\frac{1}{\pi}\frac{{\frac{1}{2}\Gamma_i}}{(t-t_i)^2+(\frac{1}{2}\Gamma_i)^2}}+A^{inc}\phi(t)
\label{eqn:fit}.
\end{equation}

% \begin{eqnarray}
% C(t) = C_{ucn}+ \hspace{ 2 in} \nonumber \\
% \{ \Sigma_{i=1}^{5}{A^{coh}_i\frac{1}{\pi}\frac{{\frac{1}{2}\Gamma_i}}{(t-t_i)^2+(\frac{1}{2}\Gamma_i)^2}}+A^{inc}\phi(t)+C_B \}.
%\label{eqn:fit}
%\end{eqnarray}

\noindent Due to the long propagation time, the count rate of UCN, $C_{ucn}$, should be flat throughout the time channels. On the other hand, the CN background $B_{CN}(t)$ has detailed, time-dependent features due to both coherent and incoherent scattering off the UCN guides. It can be modeled as the sum of distinct Bragg peaks (the five most prominent ones are included) and a continuous background due to diffuse scattering. The time-independent background $C_B$ is introduced to account for the cosmic ray background and the detector electronic noise.
In the analysis, the first 10 channels are cut due to the contamination from the prompt radiation created by the proton pulse on target.  Occasionally, the proton pulse was observed to arrive earlier than the t$_0$ trigger, so we also must cut the final 2 channels. 
The positions of the diffraction Bragg peaks are summarized in Table \ref{tbl:bragg}. The peaks were fitted using Lorentzian functions with different width $\Gamma_i$ and coherent amplitude $A^{coh}_i$, for $i=1-5$. The widths could be quite large due to the large angular acceptance of the UCN guide geometry.
The amplitudes $A^{coh}_i$, $A^{inc}$, and $C_B$ are determined simultaneously in the final stage of the fit using $\chi^2$ reduction techniques. Examples of MCS data and fitted spectra of a typical run are shown in Fig.~\ref{fig:bkgfit}.

All data from background runs with an empty cell throughout the experiment were summed up to construct a parameterized background function $B_{CN}(t)$, that is then used as a standard background function for UCN production runs.  Much care has been taken in subtracting this background function from the UCN production data. With D$_2$ (or O$_2$) solid in the cell, the primary cold neutron beam was attenuated due to elastic scattering, reducing the population of cold neutrons which could elastically scatter from the UCN guide, resulting in a reduced CN background. On the other hand, cosmic ray and electronic noise characterized by $C_B$ presumably remains constant between runs. Therefore, the background function used for UCN production runs was modified to be: 
\begin{equation}
C_{bg}(f,t)=\left((1-f)+fe^{-n\sigma(E(t))x}\right)B^{empty}_{CN}(t) + C_B,
\label{eqn:bkgmathterm}
\end{equation}
where the cold-neutron associated background function, $B^{empty}_{CN}(t)$, is derived from the empty cell data (i.e., $C_{ucn}=0$), and $f$ is the effective fill level discussed in sec.~\ref{sec:CN}. Here, $f$ quantifies the overall attenuation of the CN background through the equation above, and thus no arbitrary scale factor is required.
The intensity of the diffraction peaks changes between the filled and empty cell cases, due to the cold neutron attenuation by the scattering target, leading to a self-shielding effect. The exponential dependence characterizes the energy dependence of the cold neutron attenuation. The total cold neutron scattering cross-section, $\sigma$, was measured from the {\it in-situ} cold neutron transmission measurements (as discussed in sec.~\ref{sec:CN}), and $x$ is the length of the cell.  

The data was fit to $C(t)=C_{ucn}+C_{bg}(f,t)$ through $\chi^2$ minimization. To further increase counting statistics, different regions of interest (ROI) were separated and re-binned individually. The ROI were chosen to represent the distinct features of the data:  the last 5~ms of the frame channels were mostly UCN and cosmic background only, whereas in the range from 25-40 ms the signal consisted of UCN and incoherent elastically scattered neutrons and between 15-25~ms the signal was dominated by the diffracted neutrons. Six groups were fit separately and the final determination of the free parameters was made via a grid search algorithm that minimizes the combined $\chi^2$, defined as
\begin{equation}
\chi^2 = \sum_{i=1}^{6} \frac{[(C_{Meas} - C_{Fit})_i]^2}{\sigma_i^2}, \label{eqn:min}
\end{equation}
where $\sigma_i$ is the statistical uncertainty in the $i^{th}$ group.  Typically, the reduced $\chi^2$ was close to 1, and no larger than 1.5. The $\chi^2$ distribution is consistent with expectations. Nominal results of the fit are shown in Fig.~\ref{fig:bkgfit} for a measurement using o-D$_2$ at 4.78 K.
 
\begin{figure}[!t]
\includegraphics[width=3.5in]{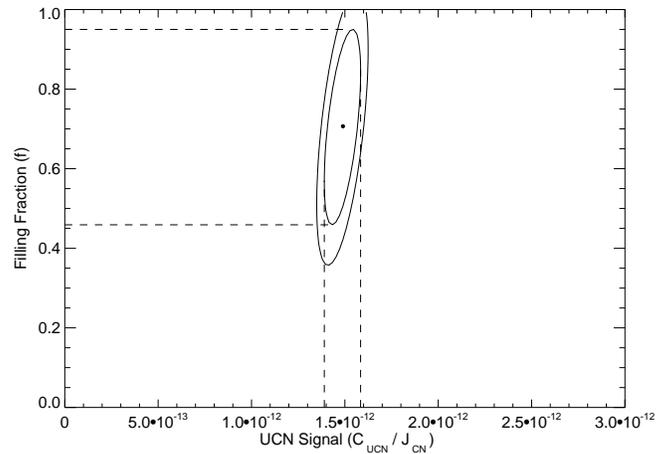}
\caption{\label{fig:chisq}A scan of $\chi^2$ for the 4.78 K measurement.  A dot indicates the location of the minimum, at reduced $\chi^2=1.26$, and the contours which enclose the 68\% and 95\% confidence levels area also drawn.  A narrow range of UCN signal is returned across a wider range of filling fractions, indicating the relative in-sensitivity to the $f$ parameter.}
\end{figure}

In the final minimization procedure, only $f$ and $C_{ucn}$ are free parameters. A typical scan of $\chi^2$ in the parameter space is shown in Fig.~\ref{fig:chisq}, which presents the analysis of a data set collected using s-D$_2$ at 4.78 K.  The result shows low correlation between $C_{ucn}$ and $f$, reflecting that the UCN count was not significantly altered by changing the background level through adjusting the fill level $f$. The minimized $\chi^2$ is 1.26 for this temperature. 
However, if one sets the filling fraction to 1, the minimized $\chi^2$ rises to 1.70, indicating a poor fit to the data.  The averaged filling fraction across all temperatures is $f=0.61 \pm 0.06$ (with the uncertainty determined from the standard deviation across the data points at different temperature), consistent with the value estimated from the total cross-section estimate and the pressure change in the gas storage volume.  Error bars on the values of $C_{ucn}$ and $f$ are set by the largest extent of the 68\% confidence level ($\Delta \chi^2=1$) in determining each free parameter. For each 1000~s run, the UCN count rate can be measured to within 8\% relative uncertainty.

%The goal is to extract the UCN counts, $C_{UCN}$, from data of the filled cell, $C(t)$, that consists of both the UCNsignal and the background, i.e.,
%\begin{equation}
%C(t)=C_{UCN}+ \alpha BG(t).
%\label{eqn:fitmathterm}
%\end{equation}
%Here $\alpha$ is a parameter to adjust the overall background level.  Before commenting on this parameter, we will discuss the fitting procedure.

%When $f$ is properly chosen, $\alpha \sim 1$, indicating we have properly applied the empty cell background to the filled cell UCN signal.  For example, in the 4.78 K experiment for $f=0.57$ as usual, the $\chi^2$ minimization returns $\alpha=0.85\pm0.25$ at 63\% confidence level and improves $\chi^2$ of the overall fit compared to $f=0$ (no attenuation).  When $f=0$, $\alpha=0.51\pm0.15$, indicating an excess of background and visa versa for the opposite case $f=1$, $\alpha=1.75^{+0.47}_{-0.51}$.

%The attenuation term impacts the UCN count rate at the level of $\sim2\%$ when $C_{UCN}$ is minimized in concert with  $\alpha$.  While this is within the present margin of error at 63\% confidence level, in an arrangement where the UCN detector's time response is not monitored such a minimization would not be possible.  This would force one to perform a more naive background subtraction and result in an overestimate of the background.  We will show a method to account for the attenuation term in the absence of the detector's time response as part of the next section.

%\section{\label{sec:results} Results}

To facilitate the comparison of UCN production between runs with different operational conditions, we define the normalized UCN signal, $S$, as the total UCN count rate ($C_{ucn}$ determined from the fitting routine) normalized to the incident cold neutron current, $J_{CN}$, and the molar amount of material under study, $N_{mol}$.
\begin{equation}
%S=\frac{1}{N_{mol}}\frac{C-B}{J_{CN}}
S=\frac{1}{N_{mol}}\frac{C_{ucn}\times 512}{J_{CN}f}.
\label{eqn:UCNSignal}
\end{equation}
Here $J_{CN}$ is the number of total neutrons passing through the cell per unit time, normalized to 100~$\mu$A of proton beam on the spallation target.  It is related to the neutron flux by $J_{CN}=A_{cell} \phi$, where $A_{cell}$ is the area of the beam incident on the production cell and $\phi$ is the cold neutron flux. $J_{CN}$ is calculated from the neutron signal measured by the cold neutron monitor M$_1$, multiplied by the ratio of the cross-sectional area of the cell and that of the monitor. The filling fraction $f$ accounts for the fact that not 100\% of the incident cold neutrons are intersecting with the target. The UCN count rate, $C_{ucn}$, has units of count rate per channel. It is thus multiplied by the total channel number, 512, which corresponds to the whole time range of 50~ms, to calculate the total count rate.  
Normalizing the UCN count rate to the neutron flux monitor is a more robust method than normalizing to the proton beam current, as this avoids introducing the additional uncertainty from the inevitable varying operational conditions of the liquid hydrogen cold neutron moderator (such as temperature, fill level and ortho/para ratio).

%The cold neutron current $J_{CN}$ is converted to actual neutron flux through the following calibration scheme. 
Since the cold neutron detectors (M$_1$ and M$_2$) are operated in the current-integrating mode, there is an additional step required to convert the detector signal $I(t)$ to the cold neutron current $J_{CN}$ passing through the cell. The total neutron current ranging from 1.5 to 20~meV is determined by: 
\begin{equation}
%J_{CN}=k \int_{19.5 \mbox{ \small ms}}^{29.3 \mbox{ \small ms} } dt I(t),
J_{CN}=k \sum_{19.5 \mbox{ \small ms}}^{29.3 \mbox{ \small ms} } I(t_i),
\label{eq:JCN}
\end{equation}
where $I(t_i)$ is the digitized voltage signal output from the neutron detector in the $i^{th}$ time channel. 
The calibration scaling factor,
\begin{equation}
%k=(8.79\pm0.12)\times10^5 \hspace{0.2 in}\frac{\mbox{neutrons}}{\mbox{Volt}},
k=(7.17\pm0.04)\times10^5 \hspace{0.2 in}\frac{\mbox{neutrons}}{\mbox{Volt}},
\end{equation}
is determined by setting the total neutron flux to the results of a recent flux measurement in FP12 \cite{seppo2008}. This updated measurement of flux using a fission chamber gives a total neutron flux of $2.0 \times 10^{7}$cm$^{-2}$s$^{-1}$ (with 100~$\mu$A of proton current) over the energy range of interest to UCN production (i.e., 1.5 to 20 meV). Note that in Eq.(\ref{eq:JCN}), only part of the measured TOF spectrum (between 19.5 to 29.3~ms (i.e., 2.7 to 6.0~meV) is used to determine the scaling factor $k$. The electronic interference from adjacent power lines (40~A) that supply the pulse tube refrigerator compressor introduced additional noise, and thus the time channels with low neutron currents are excluded from the summation.   

A simple analysis that subtracts the background using the empty cell data tends to overestimate the background for reasons previously mentioned. However, for later runs in which the nickel foil mirror replaced the bend in the neutron guide, the background was significantly reduced, and the simple background subtraction routine produced signals in reasonable agreement with the algorithm presented above. The simple subtraction approach uses the model:  
\begin{equation}
\label{eqn:simple}
S_{simple}=\frac{1}{N_{mol}}\frac{(\Sigma_i C_i-\bar{T}\Sigma_i B_i)}{J_{CN}f}
\end{equation}
where $\Sigma_i C_i$ is the total number of counts measured with a filled cell, and the background $\Sigma_i B_i$ is the total number of counts measured with an empty cell. 
%\begin{equation}
%C_N=\frac{C}{J_{CN}} \Biggl\lvert_{Filled} \mbox{ and } B=\bar T \times \frac{C}{J_{CN}}\Biggl\lvert_{empty}.
%\end{equation}
As mentioned before, the first few channels that are contaminated by the prompt pulse are cut.  
The reduction of the background from the self-shielding effect due to the presence of the UCN production medium is captured by the transmission fraction, $\bar T$, averaged over the incident cold neutron energy spectrum.  To properly account for the fact that not all cold neutrons were attenuated uniformly by a cell partially filled to fraction $f$, the transmission factor is calculated as
\begin{equation}
\bar T = (1-f)+f\times\frac{\int e^{-n\sigma(E)x}\phi(E)dE}{\int\phi(E)dE},
\end{equation}
where $\bar T = 0.651$ for the ``bend" measurement ($f=0.57$) and $\bar T = 0.691$ for the later ``mirror" measurement ($f=0.507$).

\begin{figure}[!t]
\centering
\includegraphics[width=3.5in]{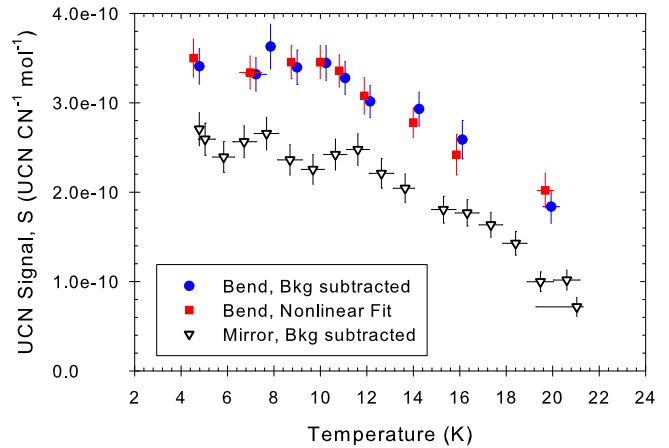}
\caption{\label{fig:result} UCN production in solid o-D$_2$ below the triple point.  The removal of the scattered cold neutron background by the fitting procedure (red squares) is commensurate with a simple background subtraction approach (blue circles). To show the difference between the two data sets, the blue circles are off-set in 
temperatures by 0.25~K. The replacement of the bend by the mirror (white triangles) reduces the UCN transmission by 24~\%.}
\end{figure}

With the proper background subtraction, we can construct a figure of merit to compare different guide geometries.
Comparison of the signal-to-noise ratio (SNR) between the mirror and the bend guides can be made using  
\begin{equation}
SNR =\frac{S}{\bar T B}.
\end{equation}
For the bend SNR is 5.6 and it improves significantly to 12.5 using the mirror.
Results using the non-linear fitting algorithm are within the error of the simple subtraction method. The experimental results from both approaches are displayed in Fig.~\ref{fig:result}.
The subtraction method is most useful when counting statistics for each individual TOF channel is so low, such that integration over all channels is required to generate a better statistical confidence in the UCN signal.  The agreement between the two approaches warrants the adoption of the simple background subtraction method for the forthcoming analysis on s-O$_2$ (despite the rather unusual background shape).

\begin{figure}
\includegraphics[width=3.5in]{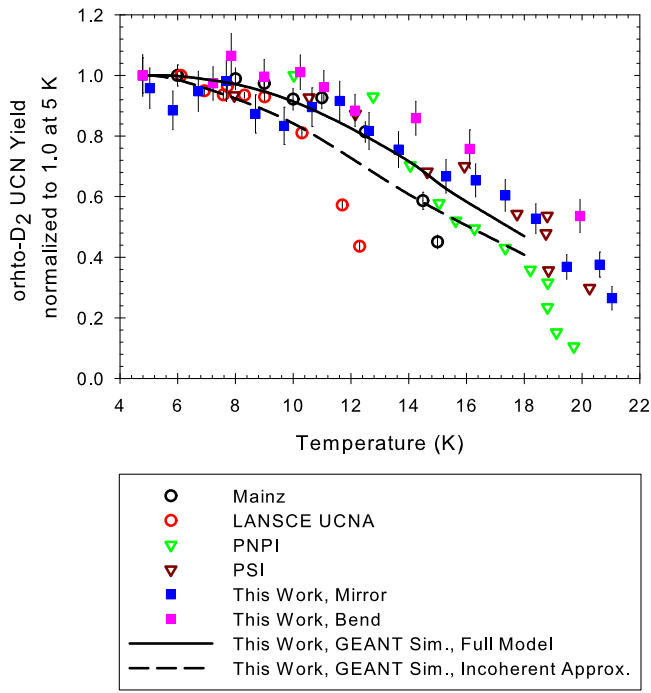}
\caption{\label{fig:comparison} The temperature dependence of this experiment compared to similar experiments (Mainz\protect\cite{lauer_thesis}, LANSCE UCNA\protect\cite{morris2002}, PNPI\protect\cite{serebrov2000a}, PSI\protect\cite{atchison2005}).  All data are scaled to unity at 5 K.}
\end{figure}

The data in Fig.~\ref{fig:result} correspond to ortho-D$_2$ in a 3.556 cm long (117.4 cc) cell with bend and mirror.
The solid was cooled over 14 hours starting from the liquid phase. In the liquid phase at 20 K, there was already a non-zero number of UCN detected well above the background level.  Solidification begins at the triple point at 18.7 K.  The observed UCN counts increased monotonically as the temperature of the production target decreased until a saturation was reached around 10 K. The enhanced UCN production was expected as the MFP of UCN increased with the reduction of the thermal phonon population, leading to a suppressed upscattering loss rate. The saturation can be understood primarily by 
the suppression of up-scattering at lower temperatures to the limit in which the escape length of UCN became comparable to the upscattering length. 
In addition, cold neutron scattering out of the cell by the UCN source was enhanced due to the increased density of the cold solid. 
A survey of results from previous experiments show similar dependence over the same temperature range\cite{morris2002,atchison2005}, with the exception of the data sets from LANL and Mainz (see Fig. \ref{fig:comparison}).

%The saturation temperature is higher than previous experiments \cite{hill2000,atchison2005} but agrees with another recent measurement at \cite{Munich} \color{red} Check reference. \color{black} The temperature dependence results from a combination of several effects - cold neutron attenuation along the axis of the cell, the suppression of up-scattering, and the UCN random walk in the cell.

\section{Discussion}
%\subsection{Monte Carlo Simulation}

We built a Monte-Carlo code on the GEANT4 framework \cite{geant4ucn}, with a complete geometry of the experiment to simulate the results using relevant UCN cross-sections.
The simulation applies the physics of UCN production in o-D$_2$, and assumes simple transport processes to track UCN from the source, through the UCN guide into the UCN detector. We simulate the expected UCN production using modest assumptions on the material properties (such as guide surface reflection specularity, UCN up-scattering cross-section in D$_2$, etc.). We then compare the measured UCN signal $S_{meas}$ to the simulated UCN signal $S_{sim}$, defined in Eq.(\ref{eqn:UCNSignal}) to fine-tune the guide parameters. Furthermore, the guide parameters were constrained using additional data on the guide transmission from the UCNA source.
We assume all guide sections have the same surface qualities, to reduced the number of free parameters in the simulation.
%The measured signal is the total detected UCN normalized to the total neutrons passing through M$_1$,
%\begin{equation}
%J_{CN}=A_{m} \delta t \phi_o,
%\end{equation}
%$A_{m}$ is the area of the neutron beam on the monitor, $\delta t$ is the measurement time, and $\phi_o$ is the total thermal and cold neutron flux over the relevant energy range.

%\begin{equation}
%\phi_o=\int_0^{\mbox{385 meV}} dE \phi(E).
%\end{equation}

\begin{figure}
\includegraphics[width=3.5in]{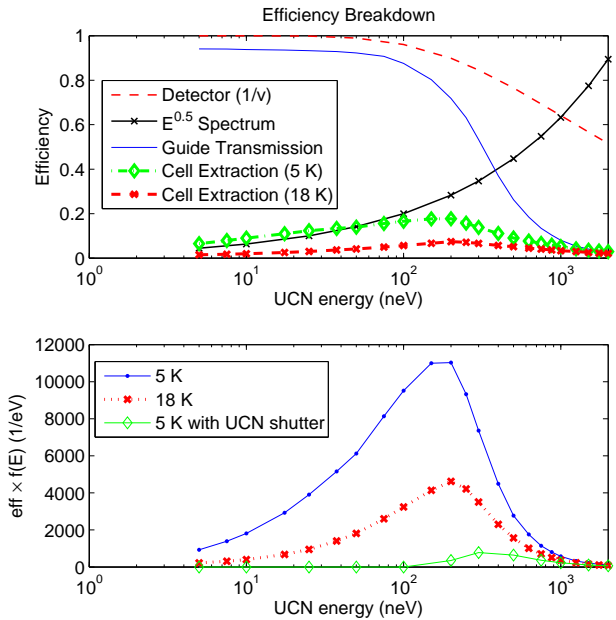}
\caption{{\bf Top:} A breakdown of efficiencies of UCN transport in the apparatus as a function of UCN energy. {\bf Bottom:} UCN transport efficiency weighted by the production spectrum $f(E)dE\propto \sqrt{E}dE$. The energy-averaged efficiency is obtained by integrating the curve up to the maximum energy of interest. The green curve is the efficiency of neutron transport through a UCN shutter used later in the experiment.}
\label{fig:TransGeant}
\end{figure}

The simulated UCN count rate $C_{sim}$ could be presented as a function of the various efficiencies through different parts of the apparatus:
\begin{equation}
C_{sim}=\varepsilon_{elec}\varepsilon_{trans}(n V_{source}\phi_o\bar{\sigma}   ),
\label{eqn:csim}
\end{equation}
where $\varepsilon_{elec}$ is the detector efficiency, $\varepsilon_{trans}$ is the transport efficiency for a UCN to travel from the source through the guide system and finally to the detector active volume, $n$ is the molecular number density, and $V_{source}$ is the volume of o-D$_2$ illuminated by the cold neutrons. 
The cross-section $\bar\sigma$ is the UCN production cross-section averaged over the incident cold neutron spectrum: 
\begin{equation}
\bar{\sigma}=\frac{1}{\phi_o}  \int_0^{\mbox{\small 1 $\mu$eV}} dE_{ucn}\int_0^{\mbox{\small 20 meV}} dE \phi(E) \sigma(E\rightarrow E_{ucn}).
\label{eq:UCNprod}
\end{equation}
%Using the measured spectrum from the neutron flux monitor, the total cross-section for UCN production averaged over the volume is
% \begin{equation}
% \bar{\sigma} = 1.48\times10^{-6} \mbox{b}.
% \end{equation}
To include the detectable VCN signal, the integration include neutron energy up to 1~$\mu$eV. 
The transmission efficiency is dependent on the neutron energy. As shown in Fig.~\ref{fig:TransGeant}, some VCN with energy larger than the Fermi potential of the UCN guide can be present among the detected signals.

%\color{red} Need a UCN transport efficiency as a function of UCN energy plot to justify the integration range: 0-1000neV. \color{black}

%To compare to the experimental results, the simulated UCN count rate can be written as
%\begin{eqnarray}
%S_{sim}&=&\frac{1}{N_{mol}}\frac{C_{sim}}{J_{CN}} \\
%&=&\varepsilon_{trans}\varepsilon_{elec}\frac{n_a \bar{\sigma} }{f A_{beam}}, \label{eqn:sim}
%\end{eqnarray}
%where the number of moles is determined from the filling fraction and the molar volume, $V_m$, as $ N_{mol}=f V_{cell}/V_m $ and $n_a$ is Avogadro's number.
%The transport efficiency is determined through simulations, the detector efficiency through detector characterization, and the averaged UCN production cross-sections using the theoretical cross-section with a full analysis of the spatial distribution of the incident cold neutron flux.

\subsection{Spatial Distribution of UCN Production}

Over the energy range relevant to UCN production, the elastic MFP of the cold neutron beam varies from 2.8 cm to 5.6 cm. Since the cell length is comparable to the elastic MFP, a large probability of elastic scattering of the incident cold neutron beam inside the cell is expected, which would significantly alter the spatial distribution of the cold neutron flux. 
This distribution is modeled using another Monte Carlo code (MCNP5 \cite{mcnp5}). The code simulates the multiple scattering process of each cold neutron and tracks the evolution of the cold neutron flux throughout the target cell. A large fraction ($\sim$ 50\%) of cold neutrons experienced at least a single scattering, and may downscatter to UCN before escaping the o-D$_2$ volume.
%We therefore must simulate the entire cold neutron path through the deuterium volume to accurately assess the expected UCN production.
%We model this effect by simulating the transport of the incident spectrum measured by M$_1$ through a volume of 19 K o-D$_2$ using MCNP5 \cite{mcnp5}.
The results of the simulation can then be compared to the measurements of the transmitted cold neutron beam collected by the detector M$_2$ (sec~\ref{sec:CN}).

To account for this spatial variation of the cold neutron flux,
a position-dependent UCN production rate, $R(z)$, is defined which varies as the distance, $z$, along the symmetry axis of the cylindrical cell. To explore the position dependence, the target cell is sub-divided into slices and the production rate is averaged over the energy spectrum of cold neutrons in each individual slice, i.e.,
\begin{equation}
R(z)=\frac{1}{\delta z} \frac{\int_{z}^{z+d_z} dz \int_o^\infty dE \sigma(E\rightarrow E_{UCN})
\phi(E,\vec{r})}{\int_{0}^{L}dz \int_o^\infty\phi(E,\vec{r})}.
\label{eqn:R}
\end{equation}

\noindent This calculation includes a divergence of the neutron beam of 0.5$^{\circ}$ at the guide exit (a result derived from another simulation using VITESS \cite{vitess}). The UCN production cross-section, $\sigma(E\rightarrow E_{UCN})$, used in the calculation is from the work of \cite{atchison2007}, integrated over final state UCN energy up to 1~$\mu$eV. The cross-section quoted in \cite{atchison2007} was scaled to the calculation using the simple Debye model and the incoherent approximation. The production rate is summed over each volume segment $dV_z$, defined as the cross-sectional area of the cell multiplied by the thickness $dz$.
The incident neutron spectrum, $\phi(E,\vec r)$, used in the Monte Carlo is measured by M$_1$.

\begin{figure}
\centering
\includegraphics[width=3.5in]{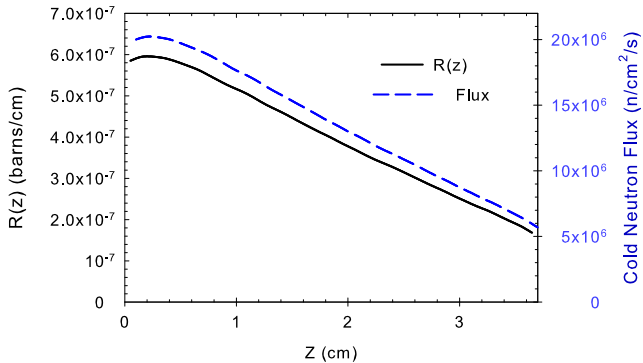}
\caption{\label{fig:sigma_z}The position dependence of the UCN production reaction rate, calculated from Eq.(\ref{eqn:R}). The integral cold neutron flux's evolution through the volume is also shown.}
\end{figure}

The position-dependent UCN production rate, $R(z)$, and the cold neutron flux, $\phi(z)$, are shown in Fig.~\ref{fig:sigma_z}, at $\delta z =$1 mm resolution.
Results of the simulation show that the cold neutron energy spectrum is not significantly changed by the presence of o-D$_2$, however, the intensity decreases along the length of the cell as more cold neutrons are elastically scattered out of the cell.
Notice that the spatial distribution does not follow a simple exponential decay, probably due to the fact that the cell is of finite volume. Finally, the cold neutron flux peaks at a few mm into the cell because of the locally enhanced cold neutron population through accumulation of elastically scattered neutrons and the increase of the neutron flux in the adjacent volume slices around the scattering site.
This spatially-dependent production rate function is then used to create the UCN source term in the GEANT4 simulation.

\subsection{\label{sec:MC}UCN Transport Efficiency}
%The transport efficiency of UCN from the cell through the guide to the detector, $\varepsilon_{trans}$, is determined by the properties of the constituent materials that UCN would interact. When traversing a material, UCN could be up-scattered CN by a material to higher energies, nuclear absorption, as well the random walk of UCN due to elastic scattering.
%One must characterize the probabilities for a UCN to absorb onto a material upon incidence, transmit into it, or reflect from the surface. Furthermore, reflection from the surface can be specular or diffusive. When traversing a material, one must also consider the up-scattering of UCN by a material to higher energies, nuclear absorption, as well the random walk of UCN due to elastic scattering.

%In the simulation, we have implemented the following physics:
%If the UCN is traversing a material medium, the transmission efficiency is governed by the nulcear absorption, incoherent elastic scattering, and upscattering cross-sections.  There may also be additional elastic scattering due to density fluctuations in the material. When transport through the UCN guide, the transport efficiency can be simulated by specifying the following material properties, such as the Fermi potential, probability of loss per collision, and the probability to scatter diffusively upon reflection.
%The gravitational deceleration is also included in the simulation.

The total transport efficiency of UCN through the apparatus can be divided into three independent components: efficiency of extraction from the cell, efficiency of transport through the guide system, and detector efficiency.
The cell, cold-shield, and nickel shield lining are modeled with parameters provided by the known absorption and scattering cross-sections. The up-scattering of UCN in these materials is assumed negligible due to their low temperatures and small thicknesses. The D$_2$ volume is then the greatest source of loss. While the scattering and absorption cross-sections are well known, the up-scattering cross-section depends on the model employed, allowing for discrimination between different models of this effect.

The diffusivity and additional elastic cross-sections are difficult to constrain. Reasonable assumptions about the magnitudes of the various parameters are made, and a comprehensive simulation is used to assess the sensitivity of the efficiency to these parameters. To this end, a UCN transport Monte Carlo based on GEANT4 is used, which includes all of the UCN transport physics. UCN are created inside the cell according to the spatial distribution above (Eq.(\ref{eqn:R})). The initial UCN spectrum is proportional to $v^2dv\sim \sqrt{E}dE$. Neutrons with energy up to 1~$\mu$eV are tracked.

The stainless steel UCN guide walls have a neutron potential energy of 189~neV, and 1\% diffusive surfaces. The loss per bounce used in our guide model is 8.5$\times 10^{-5}$. Imperfections due to welding were measured to be 2.5~cm long rings at the interface between the guides and conflat flanges, and were modeled in the Monte Carlo as 100\% diffusive regions of the guide. The foil loss was characterized in a followup experiment to measure the guide transmission using the UCNA solid-D$_2$  source. The UCN intensity in the guide system with and without the tee in place was measured, and the results are used to constrain the guide parameters. 
The diffusivity of the foil is determined to be 38\%, with which the simulation reproduces the transport reduction of 76\% when the bend is replaced by the nickel foil mirror.

Finally, the loss in passing through the detector window (a 0.51~mm Al foil) is not negligible. A transmission efficiency through the foil of 50\% for a drop of 1~m at the UCNA source was measured, similar to the drop in our system. There exists significant additional loss beyond the typical absorption and elastic scattering in Al, despite the gravitational acceleration from the 1 m vertical drop. The diffuse elastic scattering in the Al foil is tuned to agree with this measurement.  

The detailed results from the simulations reveal that about 3/4 of the UCN produced are lost before escaping the cell. The loss of UCN occurs mostly on interaction with the cell body, which is made of Al. Aluminum was used to construct the cell body for its low Fermi potential and the ease of UCN extraction, however, its large neutron absorption cross-section is a serious source of UCN loss. For a UCN with velocity of 5~m/s, only 18\% of the UCN population survives losses in the target cell and passage through the thin Al front window.
In addition to the loss in the Al cell body, 14\% of the UCN flux is lost in the solid D$_2$ at 5 K, due mostly to upscattering. Upscattering loss increases to 66\% at 18K. 
%\color{red} What energy spectrum are these numbers? \color{black}

Furthermore, the survival of UCN through the cell is dependent on the UCN energy. The maximum cell extraction efficiency is 0.15 for UCN of 200~neV.  
Integrated over the UCN spectrum up to 1~$\mu$eV, the cell extraction efficiency at 5~K is found to be 0.07, which is the fraction of UCN produced which emerge from the cell, travel through the cold shield, and enter the guide. 
The transport efficiency through the guide is also energy-dependent. With a typical cos($\theta$) angular distribution (Eq.(2.68) in \cite{golub1996}), the 
probability of transport, assuming a continuous UCN spectrum up to 1~$\mu$eV, through the guide to the detector window is 0.31. 
%In modeling the UCN guide, the loss per bounce is 8.5$\times 10^{-5}$, with a 1\% diffusivity. The Fermi potential of the guide material is 189 neV. 
Together with the transmission of 0.50 through the detector window, the $^3$He(n,p)$^3$H reaction probability of 0.86 and the full energy peak efficiency of the detector of 0.85, one estimates for an overall detection efficiency of 0.008. 
The efficiencies for UCN transport and detection are listed in Table.~\ref{tbl:est}.
%This value, along with the UCN production cross-section, is within a factor of 2 of our data.

\begin{table}[!t]
\begin{center}
\caption{\label{tbl:est}Efficiency Breakdown and UCN production Estimates. Presented numbers are spectral-weighted efficiencies and cross-sections for 
UCN with with a continuous energy spectrum up to 1~$\mu$eV to include the VCN population, and for UCN with energy up to the 300~neV in the guide, after the energy boost of 104 neV upon exiting D$_2$.} 
\begin{tabular}{|l|c|c|}
%\begin{tabular}{| p{1.5in} | p{1.5in}| p{3.85in}|}
  \hline\hline
  % after \\: \hline or \cline{col1-col2} \cline{col3-col4} ...
\textbf{Parameters} & \textbf{0-1~$\mu$eV} & \textbf{0-300~neV} \\\hline
cell extraction & 0.07 & 0.15  \\
guide transmission & 0.31 & 0.80  \\
detector window transmission & 0.50 & 0.50 \\
$^3$He(n,p)$^3$H probability & 0.86 & 0.97  \\
 $\varepsilon_{elec}$ (total energy peak) & 0.85 & 0.85 \\ \hline
total efficiency & 0.008 & 0.05  \\ \hline
$\bar{\sigma} (b) $ &1.42$\times10^{-6}$ & 1.27$\times10^{-7}$ \\ \hline
N$_{D_2}\phi_0$ (cm$^{-2}$s$^{-1}$)& \multicolumn{2}{|c|}{(44$\pm 3)\times 10^{30}$} \\ \hline \hline
  \textbf{Expected Rate} (~s$^{-1}$)& 0.49$\pm$0.03 &  0.28$\pm$0.02 \\ \hline
  \textbf{Measured Rate} (~s$^{-1}$)& \multicolumn{2}{|c|}{0.48$\pm$0.05}   \\ \hline\hline
\end{tabular}
\end{center}
\end{table}

\subsection{UCN upscattering}

The temperature dependence of UCN production originates solely within the production source. The UCN upscattering cross-section is strongly temperature dependent. The density of the target material also changes with temperature, but the dependence is much weaker by comparison. Using the upscattering cross-section calculated in our previous work\cite{liu2000}, our simulations did not reproduce the experimental data. In particular, the simulation predicts saturation at a much lower temperature, around 6 K. In this calculation, the incoherent approximation was used to treat the inelastic coherent scattering in the same fashion as the incoherent scattering.  

In evaluating the cross-section for incoherent scattering, the density of states is used to weight the contribution of the different phonon modes. 
%because of the relaxation of single particle momentum conservation, 
While the incoherent approximation works quite well to estimate the total cross-section, it might not be appropriate when evaluating cross-sections for UCN scattering, where the phase space for the process is significantly limited. 
%the energy resolution is on the order of 100 neV.  
The lack of detailed $Q$-dependence information in the incoherent approximation leads to significant errors in calculating the UCN cross-sections.

While the incoherent approximation works quite well for hydrogenous neutron moderators ($\sigma^H_{coh}$=80.27 b, $\sigma^H_{inc}$=1.7583 b), 
in the case of o-D$_2$, the contribution of coherent scattering is not small ($\sigma^D_{coh}$=5.592 b, $\sigma^D_{inc}$=2.050 b). To address this concern, a new calculation is implemented, using a full model that includes spin statistics and coherent inelastic scattering, as well as the incoherent scattering. This new calculation predicts that the revised UCN upscattering cross-section is a factor of 2-4 smaller than the calculation using the incoherent scattering. 
  
\begin{figure}
\includegraphics[width=3.5in]{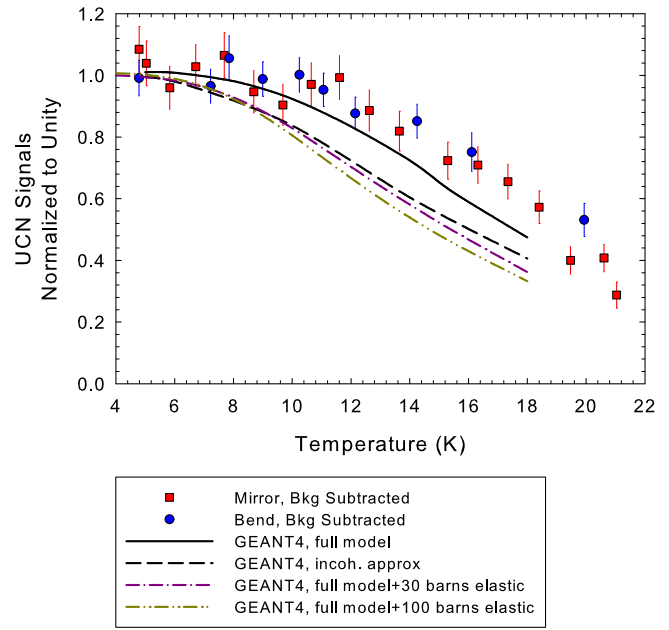}
\caption{\label{fig:geantresult}Results of o-D$_2$ UCN production experiment and GEANT4 simulation. Simulation parameters (guide efficiency, production, etc.) detailed in the text.}
\end{figure}

With an updated upscattering cross-section calculated using the full model of the dynamic structure function $S(\vec{Q},\omega)$ for o-D$_2$\footnote{to be published}, the Monte-Carlo simulation produces a temperature dependence of UCN production that agrees better with the experimental data. Both models, along with the background-subtracted data for the bend and mirror measurements, are shown in Fig.~\ref{fig:geantresult}.  
In Fig.~\ref{fig:geantresult}, to accentuate the temperature dependence, two sets of experimental data and the simulation results are scaled to agree at 4.78 K (where upscattering is small in both models) to eliminate guide and detector transmission efficiency effects.  At temperatures higher than 12 K, where the UCN upscattering cross-section is quite large ($\sim 10$ b), the incoherent model is excluded at the 1$\sigma$ level. 
Other effects that lead to increased elastic scattering inside the source, such as the presence of cracks and voids leading to inhomogeneity scattering, have been studied, however, the higher saturation temperature can only be explained by reduced upscattering cross-sections.

Increasing the elastic scattering inside the UCN production source (solid D$_2$) would lead to local trapping of UCN and thus amplify the effects of any loss mechanism inside the source. 
This effect will result in a steeper temperature dependence on the UCN production data (simulated results plotted in Fig.~\ref{fig:geantresult}).
The solid D$_2$ was solidified from liquid and cooled slowly over the duration of 10 hours without any deliberate thermal shock. 
In order to quantify the elastic scattering of UCN due to the presence of inhomogeneity, one needs to carry out measurements of UCN MFP by systematically varying the cell dimension. We have done this study on the solid O$_2$ target, but did not perform it with solid D$_2$. 
On the other hand, comparing the experimental data with GEANT4 simulations with varying degrees of UCN elastic scattering suggests that the inhomogeneity scattering of UCN inside the production target is no more than a few barns.    
This can be compared to the minimum elastic scattering cross-section, arising from incoherent elastic scattering, which is about 2 b. This result is consistent with the findings of the Los Alamos and PSI experiments on solid D$_2$ sources, which are consistent with modest variations in the effective elastic scattering cross-section for typical solid D$_2$ crystal growth.

While the results of the simulation using the updated upscattering cross-section agree quite well with our data and the data set measured independently at PSI (shown in Fig.~\ref{fig:comparison}), the data sets from the LANL and Mainz groups showed a much steeper temperature dependence for temperatures higher than 10 K. The difference comes from the different source configurations. In these two experiments, solid D$_2$ was condensed from vapor at the end of the UCN guide, which was cooled below the solidification temperature of D$_2$. The source was designed to reduce the transmission loss by eliminating a vacuum window which would contain the D$_2$. It worked quite well for low temperatures, however, at temperatures higher than 10 K, the whole UCN guide was filled with D$_2$ gas at the saturated vapor pressure. The additional upscattering from the D$_2$ vapor could be quite large, resulting in a temperature dependence steeper than the simple prediction where UCN are upscattered through single phonon absorption.

In this reported work, the essential focus is to use solid o-D$_2$ as a source to calibrate the overall efficiency of the apparatus to detect UCN. In these calibration runs, the absolute count rate of the UCN detector can be estimated using Eq.(\ref{eqn:csim}):
\begin{equation}
\mbox{Count Rate} = \varepsilon_{trans}\varepsilon_{elec}N_{D_2}\phi_o\bar\sigma, \nonumber
\end{equation}
where $N_{D_2}= n\times f V=3\times 10^{22}$ cm$^{-3}$ $\times (0.60\pm0.03) \times 117$ cm$^3$ $= (2.2\pm 0.2)\times 10^{24}$ is the total number of $D_2$ molecules in the cell. With 100~$\mu$A proton beam current on the spallation neutron target, the cold neutron flux integrated over the UCN production cross-section is
\begin{eqnarray}
\phi_0\bar\sigma &=& (2.0 \pm 0.1) \times 10^7\mbox{cm$^{-2}$s$^{-1}$}(1.42\times 10^{-6}\mbox{b})\nonumber \\
&=& (28 \pm 1.4) \times 10^{-24} \mbox{s$^{-1}$}  \nonumber
\end{eqnarray}
Here the UCN production cross-section, $\bar\sigma$, is estimated following Eq.(\ref{eq:UCNprod}) using our updated calculation, which includes the coherent scattering process. Note that the production 
cross-section reported in \cite{atchison2007} is normalized to each atom and does not include the spin statistics and the rotational form factor for molecular deuterium; the cross-section independently reported in \cite{Muller2008} should be corrected by a factor of two to properly account for the range of momentum transfer integration through $-q_{ucn}$ to $+q_{ucn}$.
With the total detection efficiency of 0.008, the expected neutron count rate is 0.49 $\pm$ 0.03~s$^{-1}$, which agrees with the measured count rate of 0.48 $\pm$ 0.05~s$^{-1}$.
%. Including the VCN contamination of 46\%, the expected count rate is 0.43, which agrees with the measured count rate of 0.43 $\pm$ 0.05 UCN~s$^{-1}$.
This confirms the general validity of our model of UCN transport.
%The discrepancy could be accounted for by adjusting some parameters that change the guide efficiency in the GEANT4 simulations, but confirms the general plausibility of our model of UCN transport .

\section{Conclusion}

An apparatus to produce and measure UCN from different converter materials has been constructed, the details of which has been presented in this paper. 
%In this UCN production apparatus, we have 40 mK temperature control over the filled cell. 
To allow for useful measurements in a pulsed neutron beam line, 
a background subtraction technique has been developed to extract UCN production data on top of a high level of background from cold neutron scattering.
Using neutrons on FP12 at LANSCE, the transport efficiency of the apparatus using o-D$_2$ solid as a calibration source has been measured.  Successful application of a Monte Carlo model which includes detailed physics of cold neutron attenuation, UCN upscattering, and inhomogeneity scattering describes the data reasonably well. With our detailed understanding of the apparatus, the analysis can be extended to UCN production in s-O$_2$, where the production rate has never been carefully measured. The results of UCN production in s-O$_2$ using this apparatus will be reported in forthcoming papers. 
Finally, the data shows an evidence for a reduced UCN upscattering cross-section as indicated by the higher saturation temperature. With an updated UCN upscattering calculated from a full coherent scattering model, the shortcomings of the widely-adopted incoherent approximation can be remedied.

\begin{acknowledgements}
We thank Phil Childress, Jim Bowers, Darren Nevitt, and Todd Sampson
in the Indiana University physics shop for their rapid, high quality fabrication of equipment used in this experiment.
We thank Bill Lozowski for his effort in preparing the nickel coated guides.
We are also grateful for the assistance provided by the Lujan Center and LANSCE.
%especially the efforts of beam line scientist Aaron Couture. 
We acknowledge Shah Vallilopy for performance of the VITESS simulation of guide divergence.
This work was supported by NSF 0457219, 0758018.
\end{acknowledgements}

\bibliography{InstPaper}

\end{document}